\documentclass[conference]{IEEEtran}

\usepackage{pgfplots}
\usepackage{graphicx}
\usepackage{array}
\usepackage{defs}
\pgfplotsset{compat=1.18}

\begin{document}
\title{Comparative Insights on Adversarial Machine Learning from Industry and Academia: A~User-Study Approach}

\author{
\IEEEauthorblockN{Vishruti Kakkad}
\IEEEauthorblockA{Carnegie Mellon University\\
vkakkad@alumni.cmu.edu}
\and
\IEEEauthorblockN{Paul Chung}
\IEEEauthorblockA{UC San Diego\\
paulc@ucsd.edu}
\and
\IEEEauthorblockN{Hanan Hibshi}
\IEEEauthorblockA{Carnegie Mellon University and \\
King Abdulaziz University\\
hhibshi@andrew.cmu.edu}
\and
\IEEEauthorblockN{Maverick Woo}
\IEEEauthorblockA{Carnegie Mellon University\\
maverick@cs.cmu.edu}
}

\maketitle
\begin{abstract}
  An exponential growth of Machine Learning (ML) and its Generative AI applications brings with it significant security challenges, often referred to as Adversarial Machine Learning (AML). In this paper, we conducted two comprehensive studies to explore the perspectives of industry professionals and students on different AML vulnerabilities and their educational strategies. In our first study, we conducted an online survey with professionals revealing a notable correlation between cybersecurity education and concern for AML threats. For our second study, we developed two Capture-the-Flag (CTF) challenges that implement Natural Language Processing and Generative AI concepts and demonstrate a poisoning attack on the training dataset. The effectiveness of these challenges was evaluated by surveying undergraduate and graduate students at Carnegie Mellon University, finding that a CTF-based approach effectively engages interest in AML threats. Based on the responses of the participants in our research, we provide detailed recommendations emphasizing the critical need for integrated security education within the ML curriculum.
\end{abstract}
\IEEEpeerreviewmaketitle

\section{Introduction}
\label{sec:intro}

The modern-day computing ecosystem consists of a large volume and variety of data. This data can be viewed in different ways by different professionals. For cybersecurity professionals, data is primarily seen through the lens of security, encompassing access levels, authorization, and data integrity. In contrast, machine learning professionals approach data as a resource, primarily for analysis, focusing on datasets and models. To explore the similarities and differences in these views, as well as to address various aspects of security, privacy, and machine learning models in today's world, we conducted comprehensive user studies. By capturing and analyzing these different perceptions, our goal was to bridge the gap between industry practices and academic research for AML threats, while developing a Capture-the-Flag (CTF) based educational approach for Adversarial Machine Learning (AML).

This paper focuses on AML, which is defined by Huang et~al.~\cite{Huang11} as the study of effective machine learning techniques against an adversarial opponent. Though Machine Learning is one of the most talked-about fields in computing, previous work has shown split opinions on testing the security of the models. In~\cite{Bieringer22}, Bieringer et~al. conducted a study of small-sized companies dealing with Machine Learning in their way of work and found that in terms of awareness of AML, some of the practitioners considered AML as important while others were either not aware or did not consider AML as a concern. They also found that ``non-AML security and AML were mingled in participants' mental models: the boundaries between the corresponding threats were often unclear.'' In another study by Mink et~al.~\cite{Mink23}, the authors interviewed $21$ participants and found that a key challenge to the awareness of AML is a lack of exposure to AML and related tools. The study also highlights the importance of educational initiatives to improve practitioners' awareness of AML threats.

In this paper, we undertake a thorough exploration of AML, examining various aspects which include theory on AML attacks and their implications, expanding our knowledge of AML attack to practical prompt-based learning models, a CTF perspective on AML, versus an industry perspective of AML. In this research, we explore different aspects of issues in Generative AI through the lens of two user studies and derive a novel educational approach for students and professionals for a defense strategy against AML attacks.

The first study involves an online survey of $12$ cybersecurity and machine learning experts where we collected and analyzed their experiences and perspectives on AML trends and defenses in the industry. In our second study, we used the concept of Generative AI to construct two CTF challenges which allowed us to evaluate two perspectives---develop our proof of concept for ML security CTF challenges for the picoCTF platform (a widely used university-hosted Capture-the-Flag educational website) and evaluate CTF as a mode of education for AML threats. In the same study, we also developed a CTF challenge where the user has to utilize Feature Collision and Convex Polytope attacks against a Generative AI chat bot with data loopback to influence the training data. 

In summary, the contributions of our work are as follows:

\begin{itemize}
    \item Our research offers deeper insights into the gap between industry and academia by analyzing participant responses from two studies and conducting comprehensive qualitative and quantitative analyses on the collected data.  
    \item We developed and evaluated our novel CTF challenges that are a proof-of-concept implementation of AML concepts, thus providing a practical and engaging learning approach for students.
    \item We have formulated detailed recommendations for integrating security education within a ML defense plan.
\end{itemize}

\section{Related Work and Background}
\label{sec:related}

We now review some background and related work on AML attacks, and then the user-centric perspective of AML.

\subsection{Taxonomy of AML Attacks}

Adversarial Machine Learning (AML) refers to the study of Machine Learning algorithms that can resist adversarial attacks that can manipulate the classification or regression algorithms which could affect the overall performance and accuracy. We provide an overview of AML attacks and the associated taxonomy, drawing key insights from the seminal literature in this field. 

Huang et~al.~\cite{Huang11} categorize AML attacks based on three properties: 
\begin{enumerate}
    \item \textbf{Influence}: This defines the attacker's ability to manipulate the training data. Causative attacks allow for direct influence, while exploratory attacks target the evaluation data. 
    \item \textbf{Security violation}: AML attacks are classified further based on the nature of the security breach. Specifically, an AML attack can be an Integrity attack where the aim of the attack is to get false positives out of the model, an Availability attack where the aim is to disrupt a system's functionality due to a large number of misclassifications, or a Privacy attack where an attacker tries to steal information about either the training data or the learning algorithm used. 
    \item \textbf{Specificity}: This dimension characterizes the level of targeting. Targeted attacks focus on manipulating specific data points, whereas indiscriminate attacks aim for a wider degradation of model performance.
\end{enumerate}

In contrast, Tabassi et~al.~\cite{Tabassi19} define the attacks based on the phase being attacked, i.e., Training vs.\@ Testing. In training phase attacks, attackers tamper with the training data to alter the model itself. Data Access Attacks are one of the training phase attacks where an attacker tries to access the confidential training set to create a substitute model for replication. Another type of AML attack in the training phase is Poisoning Attacks~\cite{Biggio2012} where the training data or the model itself are compromised by feeding the model poisoned data. Lin et~al.~\cite{Lin21} further define different types of Poisoning Attacks:
\begin{enumerate}
    \item \textbf{Label Flipping Attacks}: The labels of a model are changed in a targeted or a random way in order to trigger the misclassification of an input.
    \item \textbf{Clean Label Data Poisoning attacks}: The attackers try to inject inputs which seem legitimate data points and correct labels such that specific inputs later are misclassified by the model. One type of such attacks is \textbf{Feature Collision}, where attackers try to find a data point that is similar to the target instance in the model feature space and trick the model into misclassifying the target instance. Another type is \textbf{Convex Polytope} attacks, which consists of crafting poisonous data points in a black-box setting such that for a genuine input, the model misclassifies that instance. 
    \item \textbf{Backdoor Attacks}: Attackers try to embed a hidden backdoor in the model which allows an attacker to control the output of the model on specific inputs.   
\end{enumerate}

In regard to the testing phase, the following types of attack are defined:

\begin{itemize}
    \item \textbf{Evasion Attacks}: These attacks manipulate input data to create perturbations that are often imperceptible to humans but significantly alter the model's output.
    \item \textbf{Oracle Attacks}: In this type of attack, adversaries exploit the input-output pairings to gather information on the model like the parameters (Extraction attack), reconstruct training data (Inversion attacks) and determine in a brute force manner whether any of the supplied points were part of training dataset (Membership Inference Attack).
\end{itemize}

For the scope of this paper, we have explored training phase attacks in detail for our CTF challenge development. 

\subsection{User-Centric Perspective of AML}

In the evolving field of Machine Learning, despite the strides made in Generative AI, there is still a huge gap in understanding its implications in terms of Adversarial Machine Learning. In such a scenario, both cybersecurity and Machine Learning professionals find themselves unarmed with the necessary tools or knowledge to conduct comprehensive penetration tests on existing AI frameworks. Gupta et~al.~\cite{Gupta23} described a prompt injection attack in the ChatGPT model where the model is fed a legitimate instruction prompt and malicious user input, and the output contains confidential information about either the model or the training data. In the case of Generative AI models, prior works such as~\cite{Gupta2023, Renaud2023} have shown that, using techniques such as data poisoning attacks, modern Generative AI models are vulnerable to privacy and security attacks. 

In another study, Guo et~al.~\cite{Guo21} explored the effects of Adversarial Machine Learning on social media algorithms where they found that social network models often get trained on real-time data and hence are more vulnerable to poisoning attacks. However, the existing technologies for data cleanup before training might not be enough to detect evolving attack strategies. An example is when Microsoft introduced an AI bot on Twitter for conversational analysis~\cite{Matamoros17}. The bot interacted with other Twitter accounts and gained understanding of conversational skills with humans by training on the real-time data received during the conversation. However, within $24$ hours of deployment, the bot started to show aggression with harmful tweets as other users had made such tweets towards the bot, and the bot retrained itself based on the accumulated harmful tweets. 

Although there are various AML defense strategies and related research, we believe it is imperative to explore the impact and cause of human perception on the efficacy of these strategies. Not only does this point of view help devise effective defenses, it also helps us understand the best technique of learning for the people who are actually involved in this line of work. 

A notable study by Grover et~al.~\cite{Grover23} highlights the importance of ``AI education'' in high school curriculum where the authors explore a novel curriculum of teaching cybersecurity and AI to high school students and teachers where they talk about their application---NetsBlox for high school students in the existing \mbox{K--12} cybersecurity education. After testing their framework, they found that while it was difficult to expand on this issue, responses from teachers were overall positive. This highlights a critical need to bridge the gap between AI advancements and cybersecurity education. 

While the aforementioned studies give a deeper insight into AML and the need for its education, there seems to be little light on comparison of perspectives of two pivotal entities of the field: the professionals and students in the fields of Cybersecurity and Machine Learning. Hence, our aim is to identify the current attack landscape in the industry and associated challenges faced by students and professionals in detecting and mitigating AML attacks. We also aim to explore an academic perspective by developing AML CTF challenges in order to understand the students' point of view for this new concept in cybersecurity. 

\subsection{Defenses against AML}
\label{initialdefense}

In this section, we review prior research on defensive strategies against AML threats to understand the current security framework for ML systems. Goodfellow et~al.~\cite{Goodfellow2014} theorize a ``Generative Adversarial Networks (GAN)'' framework for estimating generative models using adversarial networks, and the authors proposed to train two models: 
\begin{itemize}
    \item Generative Model ($G$): Model that generates data
    \item Discriminative Model ($D$): Model that distinguishes between real and generated data
\end{itemize}

The training process is a mini-max scheme in which $G$ tries to maximize the error rate of $D$ using back-propagation~\cite{Huang2024} with which they are able generate high-quality data samples. Anthi et~al.~\cite{Anthi2021} discuss the rise of Intrusion Detection Systems (IDS) in Industrial Systems and highlight the vulnerabilities of ML models, including the decrease in classification performance, such as Random Forests. However, the authors were also able to improve the robustness of the supervised models by retraining the models on adversarial data points.  

Moving on to AML defense models for AI and ML systems, Gu et~al.~\cite{Gu2024} outline five key considerations for Responsible Generative AI: generating truthful content, avoiding toxic content, refusing harmful instructions, not leaking training data, and ensuring that generated content is identifiable. Zhou et~al.~\cite{Zhou2022} put forth the following defense strategies:
\begin{itemize}
    \item \textbf{Adversarial Training}: Training models with adversarial examples to improve the model robustness against actual adversarial input.
    \item \textbf{Detection Mechanisms}: Filtering out adversarial examples before they can affect the model.
    \item \textbf{Robust Optimization}: Adopting techniques such as Projected Gradient Descent (PGD) and Regularization for Neural Networks. Another paper~\cite{Gupta2018} describes the usage of PGD for consistent image reconstruction where the feedback mechanism enforces a minimal error rate. A similar concept can be applied to AML threats by projecting adversarial inputs back into a constrained set, which ensures that the model is trained to resist these perturbations. In case of regularization techniques such as the ones proposed in~\cite{Xu2015} to prevent the over-fitting problem in image classification, the same concept of reducing sensitivity to input changes can be applied to adversarial input in order to make the model more robust. Another example of this is~\cite{Xu2017}, which proposes ``feature squeezing'', which is a technique that reduces the complexity of input features to detect adversarial manipulations.
\end{itemize}

All of these approaches are proven effective against Adversarial attacks, and combining these techniques together in a single ML system has been shown to produce significant empirical results by improving the robustness of deep learning models against such attacks as tested in~\cite{Zantedeschi2017}.

\section{Study 1: An Industrial Perspective}
\label{sec:study-industry}

In our first study, the focus is on gaining insights into the types of vulnerability that exist in AML. To achieve this, we conducted interviews with industry professionals, probing their concerns and perspectives on security issues within machine learning while simultaneously correlating their responses with their background. This study drew inspiration from the work of Kumar et~al.~\cite{Kumar2020}, where the authors interviewed 28 organizations and discovered that the majority of ML engineers and incident responders lacked the skills to effectively secure ML systems. Similarly, in the work of Boenisch et~al.~\cite{Boenisch2021}, after interviewing 72 participants, the authors concluded that ML practitioners have low awareness of security and privacy threats due to security being on the sidelines and functionality being given a priority.

We defined the following hypotheses for this study:

\newcommand\hypoI{The concern for AML threats amongst the participants is influenced by their background in Security, Machine Learning, and Privacy.}
\newcommand\hypoII{Participants make use of the CTF platforms in order to educate themselves about AML threats.}
\newcommand\hypoIII{Current industry security policies follow proper cyber hygiene for AML threats.}
\newcommand\hypoIV{Non-technical factors such as time, ethics, and money are important for the adoption of security practices for AML.}
\newcommand\hypoV{Technical factors such as accuracy, time, and performance of AI models are important for the adoption of security practices for AML.}

\begin{itemize}
    \item \textbf{H1}: \hypoI
    \item \textbf{H2}: \hypoII
    \item \textbf{H3}: \hypoIII
    \item \textbf{H4}: \hypoIV
    \item \textbf{H5}: \hypoV
\end{itemize}

\subsection{Online Survey Design}

\subsubsection{Ethical Considerations}
Our data collection process for our study was organized in such a way that ethical considerations were taken into account in each step of the way. Our survey script, acknowledgments, recruitment materials, and other relevant documents were approved by our institution's Institutional Review Board (IRB). The participants were compensated 5 USD for the 10-minute online survey, which is equivalent to 30 USD per hour, which is well above the minimum wage. 

\subsubsection{Recruitment}

We conducted an online survey of 12 participants who are working professionals with varied backgrounds in Security, Machine Learning, and Privacy. To streamline participant selection, we utilized two platforms: WiCyS (Women in Cybersecurity Conference) 2023~\cite{WiCyS2023} where participants were recruited through posters distributed to attendees, and an online platform connecting students of our university with their alumni where we post our recruitment text on a message board. The interested participants in both cases were asked to contact us by email. After receiving 24 inquiries collectively from both platforms, we evaluated participants' backgrounds to verify that they were above the age of 18 and that they are a working professionals in either ML or security, and selected 12 qualifying participants out of them. The survey was hosted on Qualtrics~\cite{Qualtrics} and each participant was sent a unique link to record their responses.

\subsection{Analysis}

Our survey was taken by 12 participants. About 64\% of the participants belonged to an age ranging between 25 and 34 years. The data collected was checked for possible errors, such as missing or irrelevant text entries. We also checked responses to see if any participants had zero experience in all three fields---Cybersecurity, Privacy, and Machine Learning. After finding no errors, we analyzed the study data as listed below. 

\subsubsection{Background of Participants} 

Our understanding of the correlation between the participants' background in Machine Learning, Cybersecurity and Privacy and their concern for security in Artificial Intelligence and Machine Learning led us to ask them non-personally identifiable information such as participants' age, gender, education level and relevant experience in the above-mentioned fields. Regarding the background of the participants, we analyzed two key metrics, namely ``Education'' and ``Work Experience''. The table below shows the level of education and the level of experience that correspond to the number of participants in the given category. This table (Table~\ref{tab:background}) shows a categorization as relevant to our hypothesis by aggregating some results. The more detailed breakdown is available in the Appendix.

\begin{table}[h]
  \centering
  \vspace{-1em}
  \caption{Background of participants}
  \label{tab:background}
  \vspace{-1em}
  \begin{tabular}{|c|c|c|}
    \toprule
    Category&Topic&Count(out of 12)\\
    \midrule
    Education & Security & 7\\
     & Privacy & 4\\
     & Machine Learning & 5\\
    \midrule
    Work Experience & Security & 9\\
     & Privacy & 7\\
     & Machine Learning & 10\\
  \bottomrule
\end{tabular}
\end{table}

As seen in the table, the participants in this study display a diverse background and a set of qualified skills, demonstrating proficiency in matters related to security and privacy issues in Machine Learning models. About 95\% of the participants have worked with the ML algorithms in some form---Regression models (9 out of 12), Decision trees (10 out of 12), Support Vector Machines (6 out of 12), Neural Networks (9 out of 12) and others including Clustering algorithms (1 out of 12) and reinforcement learning (1 out of 12). Based on these observations, we explore our first hypothesis:

\medskip
\noindent \textbf{H1}: \textbf{\hypoI}
\medskip

In this case, we considered the responses indicating participants' background in security (Q1 and Q4), ML (Q2 and Q5) or privacy (Q3 and Q6) vs. the responses indicating participants' concern for AML threats (Q13 and Q20) as stated in Appendix~\ref{sec:qre-1}. We use Fisher's exact test to perform the correlation analysis adopting the methodology specified in~\cite{Sukmana2017}, and calculate the p-value. From the data and calculations shown in Appendix~\ref{sec:quantitative}, we construct the following table that denotes the combination of entities as mentioned in Table~\ref{tab:background}.

\begin{table}[h]
    \centering
  \caption{Correlation of Background and Concern}
  \label{tab:backgroundvconcern}
  \vspace{-1em}
  \begin{tabular}{|c|c|c|}
    \toprule
    Questions&$\alpha$ (up to 3 digits) & Statistical Correlation\\
    \midrule
    Q1 and Q13 & 0.017 & Yes \\
    Q2 and Q13 & 0.101 & No \\
    Q3 and Q13 & 0.06 & No \\
    Q4 and Q13 & 0.035 & Yes\\
    Q6 and Q13 & 0.001 & Yes \\
    Q5 and Q13 & 0.056 & No \\
  \bottomrule
\end{tabular}
\end{table}

From the above table, we observe an ``expertise effect'' for participants having a background in education or work experience in cybersecurity since their backgrounds are correlated with their concern for AML threats, and hence are more likely to think about security in ML. However, a background in ML itself does not show any correlation with the participant's concern for security issues in ML. Lastly, an educational background in privacy is correlated with the participant's concern, whereas merely work experience is not. This result highlights the importance of security and privacy education amongst professionals, especially ML engineers.

\subsubsection{Personal Experience in the industry}

  In the studies of Hickman et~al.~\cite{Hickman2021} and Rees et~al.~\cite{Rees2023}, the authors reviewed official documents, which consisted of EU guidelines for Trustworthy AI and organizational and government documents, respectively. In both cases, the authors have concluded that while the documents aim to address the concerns in AI, due to their general nature, they lack specific legalities required in the dynamic environment of the Generative AI and ML systems, which is essential today. 
  
  In order to test this, we include some questions from the aforementioned studies in our survey, asking participants if they have encountered any security or privacy issues in their past projects. Inspired by the findings of Zhou et~al. (2022), which highlighted security vulnerabilities in Deep Neural Networks (DNN) due to adversarial examples, we conducted a study to explore security issues across various machine learning models. Participants were asked questions about different ML models and their encounters with security issues in any of the models. Those who have experience with Neural networks (8 out of 12) were asked about their experience with different neural networks and related security issues. 

Among the various neural network architectures, Convolutional Neural Networks (CNN) were the most familiar, with 7 out of 8 participants reporting having experience with them. Despite this familiarity, only 4 out of 8 participants perceived potential security issues in CNNs. In contrast, while only 1 participant had hands-on experience with Generative Adversarial Networks (GANs), a notable 4 participants identified security vulnerabilities in this model. This suggests that even limited experience with a specific neural network type can raise significant security concerns, possibly due to the inherent complexities or known vulnerabilities of that model. Expanding on this knowledge, the participants were also asked about the type of security issues they believe best describes the type of attack that could be done on neural networks. The majority (5 out of 8) of the participants believed that an attacker might be able to poison the training data by sending a bulk of malicious data to the model, while 2 out of 8 believed that an attacker might be able to modify input data and use that to force a neural network to make incorrect predictions, and also while only 1 out of 8 participants believed that an attacker might be able to reverse engineer the model itself and extract critical information. 

All participants were asked if they had encountered any vulnerabilities in the projects that they have worked on in the industry. Of the 12 participants, 5 claimed that they have indeed encountered security vulnerabilities while working on projects based on ML. 

\subsubsection{Participant's View: AML Experiences} 

In our research, we also aimed to capture diverse perspectives of the participants on the security vulnerabilities in the ML models and corresponding defense techniques. As a follow-up to the previous section, we asked the participants about the type of vulnerabilities they noticed (see Appendix~\ref{sec:codebook}) in those ML models. We divided their responses into the following categories:

\medskip
\noindent 1. \textbf{Privacy}: Lack of privacy of the data being fed into the ML model was one of the top vulnerabilities that some of the participants found. Participant B7 pointed out that there could be ``indirect access to information of private nature''. 

\medskip
\noindent 2. \textbf{Information Security}: This was a popular mention during the survey, and the participants mentioned various types of attacks. Participant B3 and B2 were concerned about the ``defense models of ML in general'' with respect to attacks on the overall system. 

\medskip
\noindent 3. \textbf{Embedded Systems and Reverse Engineering}: Participant B3 pointed out that many ML systems have some embedded and IoT components which ``could be reverse engineered'', referring to encryption in embedded systems. We found supporting evidence of this claim in~\cite{gopalakrishna2022if}, where the authors found that IoT practitioners are often willing to compromise the security of the entire system for a reduction in cost.  

\medskip
\noindent 4. \textbf{Feature Extraction}: Participant B4 expressed their concerns regarding the feature extraction phase in ML models, claiming that ``Feature extraction in a defensive security could be evaded easily''. 

From the above observations, while the number of participants is small, a substantial fraction reported security or privacy concerns in current ML practice. We treat this as exploratory evidence that motivates further investigation of organizational cyber hygiene in the AML context:

\medskip
\noindent \textbf{H3}: \textbf{\hypoIII}
\medskip

Based on participant reports in this sample, we did not find evidence supporting H3.
Moving forward, we asked participants about whether and how professionals in ML and security keep up to date with the latest security trends. 7 out of 12 stated that they actively keep themselves updated. Their primary reason to do so was professional growth in the industry and their secondary reasons were either due to company policies or personal interests in security. We also asked them about their sources to keep themselves updated. Some of them mentioned online resources such as courses related to AML or online blogs. Participant B8 mentioned ``hacker news and feedly'' and B3 mentioned ``medium blogs''. However, social media platforms such as X (formerly Twitter) and LinkedIn are the most popular platforms for participants to stay updated about recent vulnerabilities and threats in the industry. Participant B12 stated that their knowledge was enhanced by mainstream media and related threat bulletins, while Participant B4 stated that they actually followed ``online research papers from Security and Trustworthy Machine Learning (SaTML) proceedings''. Notably, none of the participants mentioned CTF platforms as a source for staying updated on AML threats and defenses. In this sample, we did \emph{not} find evidence supporting H2:

\medskip
\noindent \textbf{H2}: \textbf{\hypoII}
\medskip

We asked the participants their thoughts on the effects of adoption of security practices in ML projects, since we were interested to know the possible reason for companies not actively adopting a security model. 68\% of the participants believed that adoption of security practices could have a larger impact on the accuracy of the ML model, while 16\% suspected that it could compromise the time taken by the model and 16\% felt that it would compromise the overall performance of the model. Accuracy was rated the most important factor; these responses are consistent with H5:

\medskip
\noindent \textbf{H5}: \textbf{\hypoV}
\medskip

Secondly, we asked participants about relevance of non-technical factors into adoption of defenses for AML, where we focused on three aspects---ethics, money, and convenience. This data highlights potential non-technical trade-offs involved in adoption of AML techniques. Overall, ethics was the most valued, with a significant majority rating it as very or extremely important. Next was convenience (in the adoption of security practices in any ML model), which was rated important but less critical than ethics. Lastly, money, while being rated important by some participants, was not a necessary priority in this case. These responses are consistent with H4 in this sample:

\medskip
\noindent \textbf{H4}: \textbf{\hypoIV}
\medskip

The participants were also asked about relevance of core security concepts such as STRIDE (Spoofing, Tampering, Repudiation, Information Disclosure, Denial of Service (DoS), and Elevation of Privilege) and CIA (Confidentiality, Integrity, and Availability) while thinking about security of ML systems. In regard to CIA, confidentiality and integrity were seen as crucial factors for data being fed into ML models compared to availability. In regard to STRIDE, information disclosure and data tampering were perceived as the most critical threats to existing ML models, followed by spoofing and repudiation. While elevation of privilege and denial of service were considered important threats by the participants, they were also the ones that the majority of the participants were least concerned about. 

Lastly, we asked the participants about their professional opinions on how to enforce best practices in privacy and security hygiene. 

\medskip
\noindent 1. \textbf{Hire experts}: Some participants stated that they believed that hiring security experts would be a good way to ensure security in ML models. Participant B5 claims ``hiring attackers to attempt to break the system or try using it in unintended ways'' would be beneficial to recognize threats, while Participant B4 believes that using ``purple teaming and collaboration with security teams'' by AI/ML teams can help cover security gaps. 

\medskip
\noindent 2. \textbf{Train Employees}: A common opinion of various participants was to actually provide technical training to ML engineers and ensure that they are aware of the security vulnerabilities out there. 

\medskip
\noindent 3. \textbf{Additional Layers in ML models}: Participant B4 stated that ``instead of having singular model with a final output, layers of different models can be used to avoid single point of failure''. Participant B7 emphasized the need for software to ``check layers of a model and identify perceived threats''. Participant B11 believed that the parameters of the model itself should be protected. 

\medskip
\noindent 4. \textbf{Improved Data Processing}: Participants B8 and B9 stated that the need for training data should be filtered before starting the training phase, especially for chat bots and other generative AI applications. This includes redaction of Personally Identifiable Information (PII) and carefully vetting the data set before feeding it back to the model. One participant also stated the need to do more ``dry runs and tests on the model'' and all their versions to prevent possible privacy breaches. 

\medskip
\noindent 5. \textbf{Better Security Hygiene practices}: While Participants B1 and B12 felt that ``attacks can be avoided if the companies followed basic security hygiene practices'', Participant B10 stressed the need to encrypt data to ensure that the data is tamper proof. Some participants also suggested the usage of cryptography as the best way to ensure the privacy of data.

\subsection{Results}

This study aimed to explore the different perspectives of 12 industry professionals specializing in security, ML and privacy. Our analysis confirmed that concerns about AML threats are significantly influenced by the participants' background in security and education in privacy but not necessarily in ML. Contrary to our hypothesis, participants do not use CTF platforms for any type of AML education, and instead preferred online resources and social media. The findings also revealed a lack of AML threat infrastructure and related defenses. Both technical factors, such as accuracy, time, and performance, and non-technical factors, including ethics, convenience, and financial considerations, were crucial in the adoption of AML security practices. Recommendations from participants included hiring security experts, providing technical training for ML engineers, implementing layered models, enhancing data processing, and adhering to better security hygiene practices to strengthen AML security.

\section{Study 2: A Capture-the-Flag Perspective}
\label{sec:study-ctf}

In this section, we explore an educational view of Adversarial Machine Learning threats. As seen in our previous studies, none of the participants mentioned CTFs as a learning resource for AML threats despite it being a popular source of cybersecurity education. In our second study, we explore whether CTF-style challenges can support learning about AML threats. For this study, our team developed CTF challenges for picoCTF wherein we developed a Proof-of-Concept implementation of a poisoning attack in the form of a chat bot. While the expression ``gamification of security'' can be a controversial one, CTF (an embodiment of this expression) is becoming more popular than ever. The study of Bratus et~al.~\cite{Bratus2010} explores a ``Hacker Curriculum'' and states that giving students a ``security culture shock'' using CTFs is important to help them question everything which is the ideal security professional mindset. Nelson et~al.~\cite{Nelson2024} designed a challenge-based curriculum where they concluded that despite the students finding the course pretty arduous, most of the students said they learned something new after solving those challenges. Similarly, in the studies of Dabrowski et~al.~\cite{Dabrowski2015} and Leune et~al.~\cite{Leune2017}, the competitive gaming-based approach was not only an effective educational tool amongst students but also increased their interest in cybersecurity. These studies provided the impetus for designing our novel CTF challenges aimed at advancing education in AML. Another paper~\cite{Feffer2024}, which aims to explore the role of red teaming in assessing Generative AI vulnerabilities, states that current red teaming exercises only serve as a regulatory check rather than a concrete security solution for Generative AI threats.

\subsection{Research Questions}

Our main research goal for the second study was to understand whether CTF challenges are a good way to educate participants about the threats of Generative AI. Firstly, we studied the interaction of students with the user interface that we developed to determine the usability of the platform. Secondly, we designed questions to evaluate students' interest and knowledge in AML and Generative AI threats. This was followed by understanding the difficulty of the challenges developed and possible future improvements in the interface before their future deployment on the picoCTF platform. Finally, we gauged the usage of CTF platforms as a teaching medium for adversarial threats in ML and Generative AI. For this study, we also defined the following initial hypothesis:

\newcommand\hypoIb{There is a correlation between the educational background of the participants and their preference for CTF as a means of cybersecurity education.}
\newcommand\hypoIIb{There is a correlation between the participants' history of CTF usage and their preference for CTF for cybersecurity education.}

\begin{itemize}
    \item \textbf{H1}: \hypoIb
    \item \textbf{H2}: \hypoIIb
\end{itemize} 

\subsection{Interviews}

\subsubsection{Development of CTF Challenges}

For this study, as mentioned above, we designed two CTF challenges based on selected AML concepts. This experimental study included two sequenced tasks that were the CTF challenges developed by us and survey questions related to the participants' background and the tasks. Each CTF challenge was time bounded to 15 minutes and the survey was time bounded to 30 minutes. 

The development of these challenges was based on the following factors:

\medskip
\noindent 1. \textbf{Level of difficulty}: The overall difficulty was evaluated based on their language, concepts involved, and prior knowledge of Machine Learning required. 

\medskip
\noindent 2. \textbf{User Interface}: Designing the user interface with good usability was very important. The evaluation of different CTF platforms by Noor et~al.~\cite{Noor2018} was done based on different aspects, one of which was System Usability Scale (SUS), where the authors defined a 10-point scale metric with odd numbers denoting positively coded statements and even numbers denoting negatively coded statements. We also considered this scale while developing our final iteration of CTF challenges, aiming to avoid negative traits such as users requiring technical support, hard-to-use interfaces, or inconsistent systems, and to integrate positive aspects such as ease of use, frequent utilization, and a quick learning curve.

\medskip
\noindent 3. \textbf{Overall Time Spent}: In order to get an estimate of the amount of time spent and also time bounding this study, we assign each CTF challenge a certain time limit. 

\medskip
The CTF challenges were developed as a local web application using the Python Flask library~\cite{Relan2019}. 

\begin{figure}[h]
  \centering
  \includegraphics[width=\linewidth]{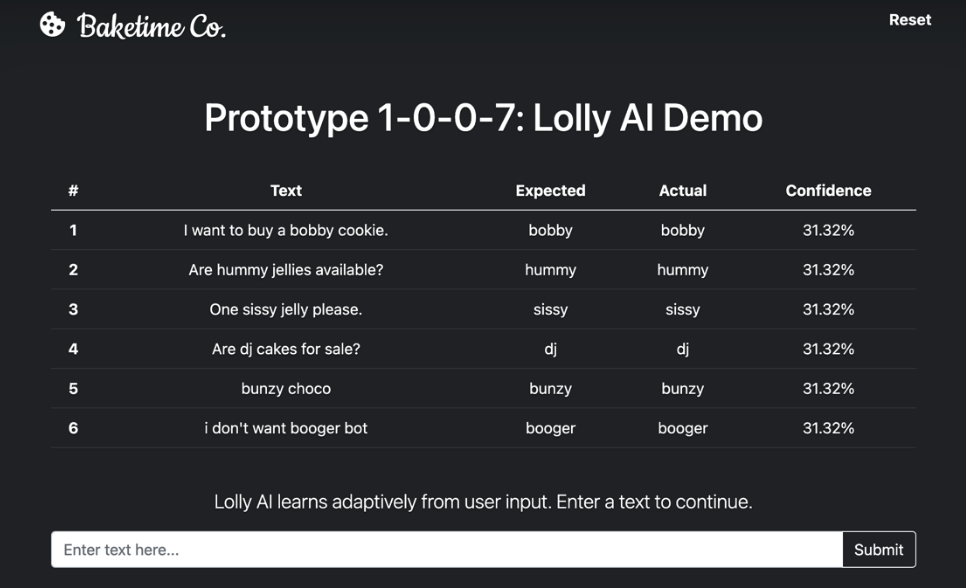}
  \caption{User Interface for Challenge 1}
  \label{fig:challenge1}
\end{figure}

\medskip
\noindent \textbf{Challenge 1 Overview}: The first challenge (LollyAI, Fig.~\ref{fig:challenge1}) was designed to demonstrate the functionality and adaptability of a self-learning machine model using Natural Language Processing (NLP) techniques~\cite{Chowdhary2020}. Participants interact with a website called LollyAI, where they are presented with test sentences for six predefined product categories. The main goal of the participant is to increase the AI's confidence rate for each category above 80\% and after achieving that goal, a flag is revealed which symbolizes success. As the participant inputs various sentences, LollyAI learns and adapts, which will in turn improve the confidence score. The difficulty level of this challenge is assumed to be ``beginner''. Participants are provided with the following interface and associated description:

\medskip
\noindent \textbf{Challenge 1 Description}: \textit{``You, as a co-employee of PicoBaker Inc., were bribed a fortune from the executives to sneak into Baketime Co. as an imposter and leak the secret recipe for their best-seller Bobby Baketime cookie. Yesterday, you were approved for your Lolly AI project that you are planning to build with Logistic Regression. Now, all you need is a larger dataset with a variety of sentences. Therefore, you decided to construct an adaptive learning system that collects users' inputs and re-trains the model. The executive board heard about your plan and became interested in seeing how the adaptive learning system will work. You decide to schedule a demo session next week. To do this, you created an interface that adaptively re-trains Lolly AI every time when an input is given. The Lolly AI demo will start with an empty dataset and low confidence rate for every category, and gradually build up as a solid model as your inputs, which is a subset of the original dataset, are given. Your current task is to rehearse the demo so that all the 6 test cases reach a confidence rate of 80\%. The demo is built so that it won't take texts shorter than 3 words without common terms, texts already in the dataset, and texts in the testing cases. Can you confirm that the demo works?''}

\medskip
\noindent \textbf{Challenge 2 Overview}: The second CTF challenge is a chatbot web application (Bake-time Lolly Chatbot), which is a website displaying different bakery products with an embedded chatbot that is used to engage users to help them with their orders. The development of this chatbot included the usage of the NLTK library~\cite{Hardeniya2015} for text processing and a machine learning model using an Support Vector Machine (SVM) with sigmoid kernel~\cite{Lin2003}. In this case, the participant's aim was to send malicious data into the chat bot in order to lower the confidence score and misclassify the category of the product, which would poison the dataset to reveal a flag. (See the Appendix for more information.) The difficulty level of this challenge is assumed to be ``advanced''. Participants are provided with the following interface and associated description:

\medskip
\noindent
\textbf{Challenge 2 Description}: \textit{``You, as a co-employee of PicoBaker Inc., were bribed a fortune from the executives of PicoBaker to sneak into Baketime Co. as an imposter and leak the secret recipe for their best-seller Bobby Baketime cookie. Some time after the approval, the imagination team of Baketime Co. successfully introduced the `Lolly' chatbot on the website. When the client interacts with the chatbot, it understands which product of Baketime Co. the client needs and takes the orders for the product on their behalf. Today, you received a message from the PicoBaker executives from a super-secure messenger, mentioning that their sales were declining during the Independence day period due to the sale of the Bobby Baketime cookie. They insisted that you should somehow sabotage the sales for the cookie using the information you know. You decide to make the Lolly chatbot from the website to falsely classify Bobby cookie to something else, so that the clients would not be able to order the Bobby cookie. Fortunately, you have access to the training dataset and you know how the system works, as it was you that created the entire pipeline. Can you accomplish the task?''}

\begin{figure}[h]
  \centering
  \includegraphics[width=\linewidth]{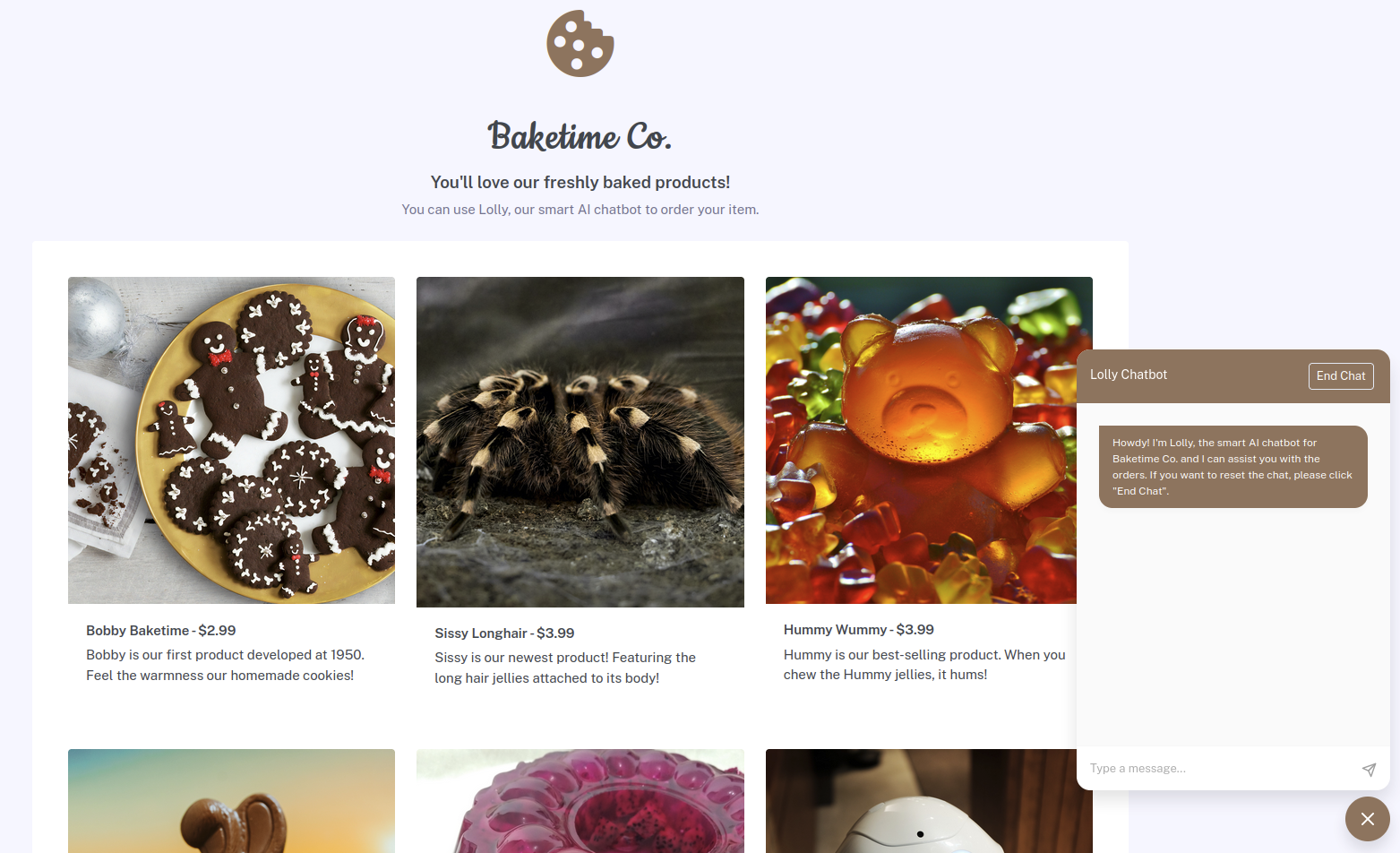}
  \caption{User Interface for Challenge 2}
  \label{fig:challenge2}
  \vspace{-1em}
\end{figure}

\subsubsection{Ethical Considerations}

Our data collection process for our study was organized in such a way that ethical considerations were taken into account at every step of the way. Our survey script, acknowledgments, posters for the study, and other relevant materials went through an approval process by the Carnegie Mellon University Institutional Review Board (IRB). The participants were compensated 50 USD for the one-hour in-person survey. No personally identifiable information or audio/video was recorded.

\subsubsection{Recruitment}

To recruit participants for this study, we designed a poster and distributed physical copies on our university campus and online on our department's Slack channel. The interested participants could contact us via the email IDs mentioned in the poster. A total of 15 participants approached us and all of them were selected as part of this study since we did not necessarily have any selection requirement except that the participants must be ``security or CTF enthusiasts''.
 
\subsection{Analysis}

\subsubsection{Background of Participants}

Of the 15 participants, 11 of them were Master's students, 3 were Bachelor students, and 1 was a Ph.D. candidate. In terms of their fields of study/experience, most of the participants (10 out of 15) have specialties (area of work/research) in Information security, some of them have backgrounds in Networks (7 out of 15) and Programming (6 out of 15) and a few in Artificial Intelligence (2 out of 15), Data Science (2 out of 15) and Psychology (1 out of 15). The participants were further asked about their background in programming since a working knowledge of programming languages aides towards a better understanding of secure coding as concluded by~\cite{Ruef2016}. While programming is not a specialty of many participants, 93\% of the participants have prior programming experience and the top languages used by them were Python, C, C++, Java and JavaScript. Approximately 93\% of the participants have prior CTF experience and the top CTF platforms that were popular amongst the participants were picoCTF (52\%), HackTheBox (21\%), TryHackMe (21\%) and VulnHub (4\%). 

\subsubsection{Participant's View: CTF Insights}

In order to evaluate our challenges, we evaluate the current state of CTF platforms for AI and ML related challenges. Participants were asked to identify topics they had learned or enhanced their knowledge by solving CTF challenges. This was done through two multiple-choice questions: one focused on general security vulnerabilities/concepts and the other on ML-specific vulnerabilities/concepts. Table~\ref{tab:ctfrole} shows the stark difference between the two based on the participant response.

\begin{table}[b]
\centering
  \caption{CTF Platform: Concepts vs. Count of participants (n=15)}
  \label{tab:ctfrole}
  \begin{tabular}{|l|l|l|l|}
    \toprule
    \textbf{Cybersecurity Threats}&\textbf{n}&\textbf{ML Threats}&\textbf{n}\\
    \midrule
    Web Exploitation&9&Reinforcement learning&1\\
    Unix/Computer Systems&4&Neural Networks&1\\
    Scripting/Shell scripting&6&Multilayer perceptrons&0\\
    Reverse Engineering&3&Hidden Markov Models&0\\
    Network Security&3&Decision trees&2\\
    Forensics&4&Clustering&1\\
    Cryptography&9&Adversarial ML&1\\
    Binary Exploitation&6&Other&1\\
    \bottomrule
  \end{tabular}
\end{table}

These results suggest that participants engaged more with general cybersecurity topics than with ML-specific threats, indicating a need for more targeted integration of ML vulnerabilities in CTF-style learning. Following this, participants were asked a series of 5-point Likert scale~\cite{Joshi2015} questions before and after attempting the challenges to evaluate the effectiveness of the CTF challenges. From the charts below (Fig.~\ref{fig:ctf-beforeperception} and Fig.~\ref{fig:ctf-afterperception}), we draw the following conclusions:

\begin{itemize}
    \item \textbf{Machine learning is an interesting subject and I would like to learn more about it (Q1)}: There was no change in the response of the participants; however, most of the participants agreed that they found ML interesting and would like to learn more about it. 
    \item \textbf{CTFs are a good platform to learn security concepts (Q2)}: Although none of the participants disagreed with the notion that CTFs are a good platform to learn security concepts, there was a slight decrease in the level of agreeability.
    \item \textbf{CTFs are a good platform to learn about non-security concepts such as Machine Learning (Q3)}: The number of participants in this case increased towards a positive side after attempting the CTF challenges, which goes to show that participants were more positive about learning AML concepts via CTF platforms.
    \item \textbf{Machine learning and security are not correlated (Q4)}: The number of participants disagreeing with the statement increased, and hence more participants believed that Machine Learning and Security are correlated. 
\end{itemize}

From these observations, participants reported increased recognition of the connection between ML and security and a more positive perception of CTF-style challenges as a way to learn AML concepts. Additionally, participants maintained a high interest in ML and acknowledged the relevance of CTFs for security education, though their enthusiasm slightly declined.

\begin{figure}
    \centering
    \begin{tikzpicture}
        \begin{axis}[
            ybar,
            bar width=0.2cm,
            enlarge x limits=0.15,
            legend style={
                at={(0.5,1.5)}, %
                anchor=north, %
                legend columns=2, %
                /tikz/every even column/.append style={column sep=0.1cm}, %
                font=\tiny
            },
            ylabel={Count},
            xlabel={Questions},
            symbolic x coords={Q1, Q2, Q3, Q4},
            xtick=data,
            nodes near coords,
            ymin=0,
            ymax=12,
            width=0.5\textwidth, %
            height=5.2cm, %
            ]
            \addplot coordinates {(Q1, 0) (Q2, 0) (Q3, 0) (Q4, 6)};
            \addplot coordinates {(Q1, 0) (Q2, 0) (Q3, 3) (Q4, 3)};
            \addplot coordinates {(Q1, 3) (Q2, 0) (Q3, 7) (Q4, 2)};
            \addplot coordinates {(Q1, 7) (Q2, 4) (Q3, 2) (Q4, 3)};
            \addplot coordinates {(Q1, 5) (Q2, 11) (Q3, 3) (Q4, 1)};
            
            \legend{Strongly Disagree, Somewhat Disagree, Neither agree nor disagree, Somewhat Agree, Strongly Agree}
        \end{axis}
    \end{tikzpicture}
    \vspace{-1em}
    \caption{Perception of participants before attempting CTF challenges}
    \label{fig:ctf-beforeperception}

\end{figure}

\begin{figure}
    \centering
    \begin{tikzpicture}
        \begin{axis}[
            ybar,
            bar width=0.2cm,
            enlarge x limits=0.15,
            ylabel={Count},
            xlabel={Questions},
            symbolic x coords={Q1, Q2, Q3, Q4},
            xtick=data,
            nodes near coords,
            ymin=0,
            ymax=12,
            width=0.5\textwidth, %
            height=5cm, %
            ]
            \addplot coordinates {(Q1, 0) (Q2, 0) (Q3, 0) (Q4, 4)};
            \addplot coordinates {(Q1, 0) (Q2, 0) (Q3, 2) (Q4, 5)};
            \addplot coordinates {(Q1, 3) (Q2, 1) (Q3, 5) (Q4, 2)};
            \addplot coordinates {(Q1, 7) (Q2, 7) (Q3, 5) (Q4, 1)};
            \addplot coordinates {(Q1, 5) (Q2, 7) (Q3, 3) (Q4, 1)};
            
        \end{axis}
    \end{tikzpicture}
    \vspace{-1em}
    \caption{Perception of participants after attempting CTF challenges}
    \label{fig:ctf-afterperception}
\end{figure}

Participants were also asked questions about the level of difficulty and the overall design of the CTF challenges. When asked about the adequacy of the 15-minute time limit for each challenge, 73\% of the participants found the time sufficient for the first challenge and 53\% for the second challenge. Those who felt that the time was insufficient suggested an average ideal duration of 25 minutes for Challenge 1 and 30 minutes for Challenge 2. Secondly, participants were asked some 5-point Likert scale questions~\cite{Joshi2015} about different components of the CTF challenges. Participants rated the difficulty of the challenges on a scale from 1 to 5, where 1 represented very difficult and 5 represented very easy. From the chart below, we conclude that most participants found the first challenge between neutral and easy and the second challenge comparatively difficult in every aspect (F1: The solution, F2: The overall challenge, F3: Language of the question/CTF description, F4: Formulating the approach). This generally aligns with our assumptions. In total, 12 out of 15 participants were able to solve the first challenge and find the flag, but none of the participants were able to solve the second challenge. 

\begin{figure}[h!]
    \begin{tikzpicture}
        \begin{axis}[
            ybar,
            bar width=0.5cm,
            width=0.5\textwidth,
            height=5cm,
            ymin=0, ymax=5,
            xlabel={Field},
            ylabel={Mean of Likert Scale},
            symbolic x coords={F1, F2, F3, F4},
            xtick=data,
            nodes near coords,
            legend style={at={(0.5,-0.2)}, anchor=north,legend columns=-1},
            enlarge x limits={abs=0.75cm}
        ]
        \addplot coordinates {(F1,3.93) (F2,3.64) (F3,3.80) (F4,3.60)};
        \addplot coordinates {(F1,2.21) (F2,2.29) (F3,3.40) (F4,2.80)};
        \legend{Challenge 1, Challenge 2}
        \end{axis}
    \end{tikzpicture}
    \caption{Difficulty of the Challenges}
    \label{fig:challenge_difficulty}
\end{figure}

For statistical analysis, we used Fisher's exact test to assess the correlation between the participant's background and their preference for CTF as a means of learning AML, which is our first hypothesis. For this, we selected Q2 and Q28 from the survey (see Appendix~\ref{sec:quantitative}) where Q2 represents the educational background of the participants and Q28 represents their preference for CTF. After converting the 5-point Likert responses (1: strongly disagree, 2: disagree, 3:neutral, 4:agree and 5: strongly agree) to a 3-category response (Yes, No and Neutral) for a clearer view, Fishers Exact Test (see Appendix~\ref{sec:quantitative}) yielded an $\alpha$ value of $0.006$. Since the value of $\alpha$ is less than $0.05$, we find evidence consistent with H1: 

\textbf{H1}: \textbf{\hypoIb}

For H2, we analyzed the responses to Q7 and Q28 (see Appendix~\ref{sec:quantitative}), where Q7 represents the background of the CTF of the participants and Q28 represents their preference for CTFs. Using Fisher's Exact Test, we obtained $\alpha = 0.6$, which indicates that we did \emph{not} find evidence supporting an association between prior CTF usage and preference for CTFs as an educational tool in this sample for H2:

\textbf{H2}: \textbf{\hypoIIb}

\subsection{Results}

This study aimed to evaluate the effectiveness of CTF challenges for AML education, revealing key qualitative and statistical insights. Our participants found our first challenge easier than the second, suggesting a need to balance difficulty and time allocation in both challenges. Participant engagement was comparatively higher for generic cybersecurity threats compared to newer ML-based threats, indicating a need to introduce AML into CTF platforms. In terms of the CTF challenges pertaining to this study, participants showed increased positivity towards learning AML concepts and recognized a correlation between ML and security despite a slight decline in enthusiasm for CTFs as a learning platform in general. We also found that while there is a correlation between the participant's background in security and their preference of using CTFs as a learning mode for AML, there was no correlation between the participant's past background in usage of CTFs and their usage of CTFs to learn AML concepts. This implies that regular users of CTF platforms do not use CTFs for AML education.

\section{Conclusion}
\label{sec:recommendations}

In this section, we draw insights from our user studies and existing literature and propose the following recommendations to bridge the gap between the current industry professionals in ML and academia by developing a more comprehensive approach to defend against AML threats:

\noindent 1. \textbf{Adversarial Training}: Incorporating adversarial sample data points during the training phase of an ML model was emphasized by our participants. Extensive test case scenarios should be created to achieve optimal balance between the accuracy and robustness of a model. 

\noindent 2. \textbf{Continuous Training and Education}: We should increase training for industry professionals and students in security and ML. From our study, we conclude that while there are academic developments in AML threats and defenses, our participants believe that there is no focus on hands-on training for such threats. Instead of relying on professionals to be proactive and use online resources to stay up to date, schools and industries should implement training programs and encourage the use and development of formal learning resources for the same. 

\noindent 3. \textbf{Data Preprocessing and Filtering}: There is also a need for preprocessing and filtering the training data to prevent adversarial manipulation, as pointed out by the participants in our study. We recommend implementing rigorous data vetting and sanitizing, including manual reviews and duplicate removal, and applying data augmentation techniques such as rotations, translations, synthetic data generation using GANs, and noise additions to enhance model resilience.

\noindent 4. \textbf{Integrate Standard Security Principles}: ML systems should also implement the basic security principles such as the CIA triad and STRIDE in their architecture. Data Privacy should be given a higher priority, and the system should not be storing any kind of Personally Identifiable Information (PII) in order to protect data integrity. Stringent access control should be applied as well. 

\noindent 5. \textbf{Ethical and Non-Technical Considerations}: Defending against AML attacks as we see in our first study also involves taking into account ethical guidelines and non-technical factors such as convenience and financial implications. The practice of involving a diverse group of decision makers and experts in addition to ML engineers such as security engineers, developers, and policy makers should be promoted to gather insights and continuously improve guidelines. 

\noindent 6. \textbf{CTF-based Learning Approach}: From our second study, we conclude that participants (even active CTF users) do not use CTFs to learn about AML concepts and most CTF platforms out there do not have related challenges.  Hence, we propose active development and formalization of CTFs related to ML concepts and its integration in the industry and academia to not only educate people about AML threats but also develop an integrated culture of learning security threats in ML with other fields of Cybersecurity. 

\section{Future Work}
\label{sec:future}

As we saw in Sections~\ref{sec:study-industry} and~\ref{sec:study-ctf}, current tools for AML education are insufficient for professionals and students, resulting in a need for some resources that provide hands-on training. We proposed a novel CTF educational model tailored for AML education while incorporating insights from our study. Considering the insights from our study, we suggest the following revisions to our existing CTF implementation for the future:

\noindent 1. \textbf{Difficulty Levels}: CTF challenges should be designed with an incremental difficulty level as the user progresses through them. We propose the following concepts for each difficulty level:
\begin{itemize}
    \item Beginner: Basic challenges involving understanding of ML models and concepts like confidence score, classification, regression, etc.\@ using popular ML Python libraries~\cite{Gevorkyan2019}.
    \item Intermediate: Creating proof-of-concept implementations of popular AML attacks such as Data Poisoning attacks and Label Flipping attacks.
    \item Advanced: Exploration of defense techniques such as Adversarial Training, PGD, and usage of GANs.
\end{itemize}

\noindent 2. \textbf{User Interface Improvements}: While many participants liked our unique user interface, some faced a learning curve despite the challenge descriptions offered to them. Hence, we want to utilize feedback loop for the next version of these challenges so that users can report issues and suggest improvements. For first-time users who are new to the concept of Adversarial Machine Learning, we want to introduce pop-up hints or links to relevant articles. 

\noindent 3. \textbf{Incorporating Realistic Scenarios}: While our challenges were rated ``fun'' by participants, some participants found the CTF scenarios to be slightly unrelatable. Hence, we plan to incorporate real-world AML scenarios such as~\cite{Matamoros17} to demonstrate the impact of adversarial attacks. We would also like to create some cross-disciplinary components to some challenges that require the knowledge of both ML and cybersecurity, encouraging participants to think holistically instead of merely thinking about AML threats as a separate branch. 

\section{Limitations}
\label{sec:limitations}

We acknowledge that there were some limitations to our study. For our first study, our participant pool was relatively small, and while the participants worked at different companies, the data size is not enough to make a generalization for the whole industry. A substantial proportion of the participants were alumni or students who studied at Carnegie Mellon University. This skew in the demographic of participants can lead to a lack of diversity, potentially introducing bias that might affect the overall applicability of our findings.

\bibliographystyle{IEEETran}
\bibliography{references}

\appendix
\section{Appendix}

\subsection{Questionnaire for first study}
\label{sec:qre-1}

\begin{enumerate}
    \item[Q1.] How much experience do you have in the field of cybersecurity?
        (No experience
        / Roughly one year
        / One to less than five years
        / Five to less than ten years
        / More than ten years)

    \item[Q2.] How much experience do you have in the field of Machine Learning?
        (No experience
        / Roughly one year
        / One to less than five years
        / Five to less than ten years
        / More than ten years)

    \item[Q3.] How much experience do you have in the field of Privacy?
        (No experience
        / Roughly one year
        / One to less than five years
        / Five to less than ten years
        / More than ten years)

    \item[Q4.] Do you have an educational background in the field of cybersecurity? If yes, which of the following best describes your background?
    \begin{itemize}
        \item No, I do not have an educational background in cybersecurity
        \item Bachelor's in cybersecurity
        \item Bachelor's in another field and minor in cybersecurity
        \item Master's in cybersecurity
        \item Master's in another field and minor in cybersecurity
        \item Ph.D. in cybersecurity or a related field
        \item I have taken some courses on cybersecurity either in my Bachelor, Master's, or online courses.
    \end{itemize}

    \item[Q5.] Do you have an educational background in Machine Learning? If yes, which of the following best describes your background?
    \begin{itemize}
        \item No, I do not have an educational background in ML
        \item Bachelor's in ML
        \item Bachelor's in another field and minor in ML
        \item Master's in ML
        \item Master's in another field and minor in ML
        \item Ph.D. in ML or a related field
        \item I have taken some courses on ML either in my Bachelor, Master's, or online courses.
    \end{itemize}

    \item[Q6.] Do you have an educational background in the field of privacy? If yes, which of the following best describes your background?
    \begin{itemize}
        \item No, I do not have an educational background in privacy
        \item Bachelor's in privacy
        \item Bachelor's in another field and minor in privacy
        \item Master's in privacy
        \item Master's in another field and minor in privacy
        \item Ph.D. in privacy or a related field
        \item I have taken some courses on privacy either in my Bachelor, Master's, or online courses.
    \end{itemize}

    \item[Q7.] Circle the ML models that you have previously worked with at a code level or in a project:
        Regression
        / Decision Tree
        / Support Vector Machines (SVMs)
        / Neural Networks
        / Other (please specify)

    \item[Q8.] If Neural Networks are selected, select the neural networks that you have worked with:
    \begin{itemize}
        \item Multi-layer perceptron
        \item Convolutional Neural Networks (CNN)
        \item Recurrent Neural Networks (RNN)
        \item Long short-term memory network (LSTM)
        \item Generative Adversarial Network (GAN)
        \item Other: \_\_\_\_\_\_\_\_\_\_
    \end{itemize}

    \item[Q9.] If Neural Networks are selected in Q7, select the neural networks that you believe have/might have security issues/ vulnerabilities:
    \begin{itemize}
        \item Multi-layer perceptron
        \item Convolutional Neural Networks (CNN)
        \item Recurrent Neural Networks (RNN)
        \item Long short-term memory network (LSTM)
        \item Generative Adversarial Network (GAN)
        \item Other: \_\_\_\_\_\_\_\_\_\_
    \end{itemize}

    \item[Q10.] In your opinion, how relevant is security in the field of Machine Learning?
    \begin{itemize}
        \item Extremely relevant - there is a lot of interdependence
        \item Moderately relevant - there is some interdependence
        \item Relevant, but there is no interdependence
        \item Not Relevant - there is no interdependence
        \item I have no opinion regarding this
    \end{itemize}

    \item[Q11.] In your opinion, how relevant is privacy in the field of Machine Learning?
    \begin{itemize}
        \item Extremely relevant - there is a lot of interdependence
        \item Moderately relevant - there is some interdependence
        \item Relevant, but there is no interdependence
        \item Not Relevant - there is no interdependence
        \item I have no opinion regarding this
    \end{itemize}

    \item[Q12.] Have you encountered a security issue related to Machine learning during a project? (Yes/No)

    \item[Q13.] ``Machine learning is essential in today's computer ecosystem.'' In your opinion, rate the statement on a scale of 1 to 5, where 1 = strongly disagree, 2 = disagree, 3 = neutral, 4 = agree and 5 = strongly agree.

    \item[Q14.] If yes, in your opinion, what was the severity of the security issue concerning the whole project?
        (Very low / 
        Low /
        Moderate /
        High /
        Very high)

    \item[Q15.] If yes, can you share the type of vulnerability found? Feel free not to disclose private information about the vendor, organization, etc.

    \item[Q16.] According to you, which of the following could be true about any Neural Network:
    \begin{itemize}
        \item An attacker might be able to modify input data and use that to make the neural network make incorrect predictions
        \item An attacker might be able to poison the training data by sending a bulk of malicious data to the dataset
        \item An attacker might be able to reverse engineer a neural network and extract information
        \item An attacker might be able to violate privacy regulations by finding a fault in the training data set
        \item None of the above
    \end{itemize}

    \item[Q17.] Do you keep yourself updated with security threats and vulnerabilities in Machine learning? (Yes/No)

    \item[Q18.] If yes, select a reason why you keep yourself updated:
        Professional growth /
        Company Policies /
        Personal interest in security /
        Personal interest in ML /
        Other (please specify)

    \item[Q19.] If yes, describe in one sentence how you keep yourself updated with the same.

    \item[Q20.] If Machine Learning projects consist of a security component, in your opinion, rate the impact that such a policy can have on the following features of the ML model used:
        (i) Accuracy,
        (ii) Time,
        (iii) Performance.

    \item[Q21.] Cybersecurity consists of a concept known as the CIA triad, which consists of three fundamental principles: C = Confidentiality = Protection of sensitive information and prevent unauthorized access; I = Integrity = Accuracy of the data and prevention of modification of data and A = Availability = Information access is available whenever it is needed. On a scale of 1 to 5, how concerned would you be about possible security issues in Machine Learning to each of the components of the CIA triad? (1 - Not concerned at all, 2 - Slightly concerned, 3 - Moderately concerned, 4 - Very concerned, 5 - Extremely concerned)

    \item[Q22.] Another concept in cybersecurity is the STRIDE model. The STRIDE model consists of the following - S: Spoofing of user identity where a user pretends to be a victim during an attack, T: Tampering, R: repudiation refers to the situation where a user denies having acted; I: Information disclosure such as privacy breach or data leak, D: Denial of Service and E: Elevation of Privilege. On a scale of 1 to 5, how concerned would you be about possible security issues in Machine Learning to each component of the STRIDE model? (1 - Not concerned at all, 2 - Slightly concerned, 3 - Moderately concerned, 4 - Very concerned, 5 - Extremely concerned)

    \item[Q23.] In your opinion, how can companies protect the security of the machine learning models against cyberattacks?

    \item[Q24.] In your opinion, how can companies protect the data privacy of the machine learning models against cyberattacks?

    \item[Q25.] In your opinion, how important are the following factors when considering security in Machine Learning (1 - Not at all important, 2 - Slightly important, 3 - Moderately important, 4 - Very important, 5 - Extremely important):
        (i) Ethics;
        (ii) Money;
        (iii) Performance;
        (iv) Convenience.

    \item[Q26.] Please select the gender that you best identify yourself with:
        Male
        / Female
        / Non-binary
        / Prefer not to say
        / Other (please specify)

    \item[Q27.] What is your age range?
        (18--24 /
        25--34 /
        35--44 /
        45--54 /
        55--64 /
        65 or older)

\end{enumerate}

\subsection{Questionnaire for second study}
\label{sec:qre-2}

This is a $60$-minute user study as part of a project at an IRB-approved academic research study related to a widely used educational CTF platform. As part of this survey, we would be collecting data based on your answers to the following questions but it will not contain any personally identifiable information that may link back to you. We will also not collect any type of other information such as audio, video, keystrokes, mouse movement, etc. that may be concerning to many people's privacy. If you have questions, please contact the corresponding author listed on the first page.

\begin{enumerate}
    \item[Q1.] What is the highest level of your degree?
        Middle school diploma or below /
        High school student /
        High school graduate, GED or equivalent /
        Some college but no degree /
        Associate, trade degree or equivalent (2 years) /
        Bachelor's degree or equivalent (4 years) /
        Master's degree /
        Ph.D. Candidate /
        Ph.D. Degree /
        Other (field)

    \item[Q2.] If your education level is higher than a high school graduate, what is your major?
    
    \item[Q3.] Do you have any prior experience in any of the fields in Computer? (Yes/No)

    \item[Q4.] What fields in CS are your specialties?
        Artificial Intelligence
        / Computer-Human Interface
        / Computer Vision
        / Computer Graphics
        / Data Science
        / Game Design
        / Information Security
        / Networks
        / Operating Systems
        / Programming Languages
        / Other (Please specify)

    \item[Q5.] Do you have any prior programming experience? (Yes/No)

    \item[Q6.] If yes, which of the following languages are you familiar with?
        C / C++;
        C\#;
        Perl;
        Java;
        JavaScript / TypeScript;
        PHP;
        Python;
        Rust;
        Swift;
        Visual Basic;
        Other (Please specify)

    \item[Q7.] Have you ever used a CTF platform? (Yes/No)

    \item[Q8.] If yes to Q7, which one(s)?
        picoCTF;
        HackTheBox;
        TryHackMe;
        VulnHub;
        HackMyVM;
        Root-me;
        Other (Please specify)

    \item[Q9.] Rate the following topics from a scale of 1 to 5 where 1 = beginner, 2 = advanced beginner, 3 = intermediate, 4 = proficient, and 5 = advanced:
        Binary Exploitation;
        Cryptography;
        Forensics;
        Web Exploitation;
        Reverse Engineering;
        Network Security;
        Unix;
        Scripting / shell scripting;
        Computer systems

    \item[Q10.] Has CTF platform helped you in learning any of the following? (Circle.)
        Binary Exploitation;
        Cryptography;
        Forensics;
        Web Exploitation;
        Reverse Engineering;
        Network Security;
        Unix;
        Scripting / shell scripting;
        Computer systems

    \item[Q11.] Rate the following topics from a scale of 1 to 5 where 1 = beginner, 2 = advanced beginner, 3 = intermediate, 4 = proficient, and 5 = advanced:
        Neural Networks;
        Decision Trees;
        Reinforcement Learning;
        Clustering;
        Multilayer Perceptrons;
        Hidden Markov Models;
        Adversarial Machine Learning;
        Other (please mention)

    \item[Q12.] Has CTF platform helped you in learning any of the following?
        Neural Networks;
        Decision Trees;
        Reinforcement Learning;
        Clustering;
        Multilayer Perceptrons;
        Hidden Markov Models;
        Adversarial Machine Learning;
        Other (please mention)

    \item[Q13.] In your opinion, rate the following on a scale of 1 to 5 where: 1 = strongly disagree, 2 = disagree, 3 = neutral, 4 = agree, 5 = strongly agree
    \begin{itemize}
        \item Machine learning is an interesting subject and I would like to learn more about it
        \item CTFs are a good platform to learn about security concepts
        \item CTFs are a good platform to learn about non-security concepts such as Machine Learning
        \item Machine learning and security are not correlated
    \end{itemize}

    \item[Q14.] On a scale of 1 to 5: 1 - Extremely unlikely, 2 - Unlikely, 3 - Neutral, 4 - Likely, 5 - Extremely likely:
    \begin{itemize}
        \item I would use CTF platforms such as picoCTF in the future to learn more about security concepts
        \item I would use CTF platforms such as picoCTF in the future to learn more about ML concepts
    \end{itemize}
    \end{enumerate}

    \subsubsection{Challenge - 1 (15 minutes)}

    \begin{enumerate}
    \item[Q15.] Challenge - 1 questions 
    \begin{itemize}
        \item In your opinion rate the following on a scale of 1 to 5 (in terms of your understanding) where 1 - Very difficult, 2 - Difficult, 3 - Neutral, 4 - Easy, 5 - Very easy:
        \begin{itemize}
            \item The overall challenge
            \item Language of the question / CTF description
            \item The solution
            \item Formulating the approach
        \end{itemize}
        \item Was the time provided just the right amount of time?(Yes/No)
        \item If no to 2, what would be the ideal time for you (minutes):
            5 /
            10 /
            20 /
            25 /
            30 /
            More than 30
    \end{itemize}
    \end{enumerate}

    \subsubsection{Challenge - 2 (15 minutes)}

    \begin{enumerate}
    \item[Q16.] Challenge - 2 questions 
    \begin{itemize}
        \item In your opinion rate the following on a scale of 1 to 5 (in terms of your understanding) where 1 - Very difficult, 2 - Difficult, 3 - Neutral, 4 - Easy, 5 - Very easy:
        \begin{itemize}
            \item The overall challenge
            \item Language of the question / CTF description
            \item The solution
            \item Formulating the approach
        \end{itemize}
        \item Was the time provided just the right amount of time?(Yes/No)
        \item If no to 2, what would be the ideal time for you: 
            5 minutes /
            10 minutes /
            20 minutes /
            25 minutes /
            30 minutes /
            More than 30 minutes
    \end{itemize}
    \end{enumerate}

    \subsubsection{General Post-Challenge Questions}

    \begin{enumerate}
    \item[Q17.] In your opinion, rate the following on a scale of 1 to 5 where 1 = strongly disagree, 2 = disagree, 3 = neutral, 4 = agree, 5 = strongly agree
    \begin{itemize}
        \item These challenges have improved my knowledge on security
        \item These challenges have improved my knowledge on adversarial machine learning
        \item These challenges have helped me learn something new in the field of security
        \item These challenges have helped me learn something new in the field of ML
        \item These challenges have improved my programming skills
    \end{itemize}

    \item[Q18.] In your opinion, rate the following on a scale of 1 to 5 where 1 = strongly disagree, 2 = disagree, 3 = neutral, 4 = agree, 5 = strongly agree
    \begin{itemize}
        \item Machine learning is an interesting subject and I would like to learn more about it
        \item CTFs are a good platform to learn about security concepts
        \item CTFs are a good platform to learn about non-security concepts such as Machine Learning
        \item I would use CTF platforms such as picoCTF in the future
        \item Machine learning and security are not correlated
    \end{itemize}

    \item[Q19.] On a scale of 1 to 5: 1 - Extremely unlikely, 2 - Unlikely, 3 - Neutral, 4 - Likely, 5 - Extremely likely:
    \begin{itemize}
        \item I would use CTF platforms such as picoCTF in the future to learn more about cybersecurity concepts
        \item I would use CTF platforms such as picoCTF in the future to learn more about ML concepts
    \end{itemize}
\end{enumerate}

\subsection{Qualitative Analysis}
\label{sec:codebook}

\begin{figure}[t]
  \centering
  \includegraphics[width=0.8\linewidth]{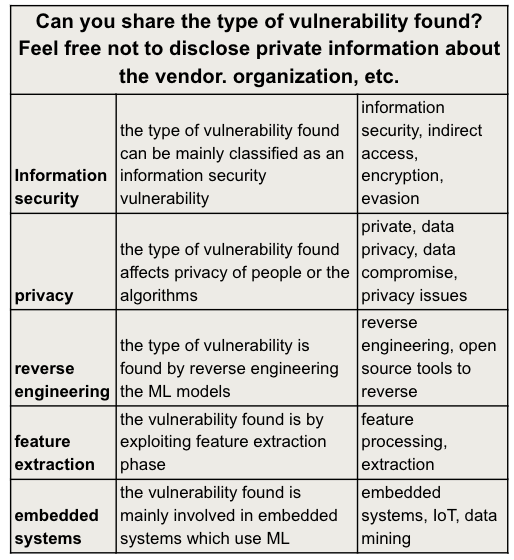}
  \vspace{-1em}
  \caption{Type of Vulnerability found in ML frameworks}
  \label{fig:a}
\end{figure}

\begin{figure}[t]
  \centering
  \includegraphics[width=0.9\linewidth]{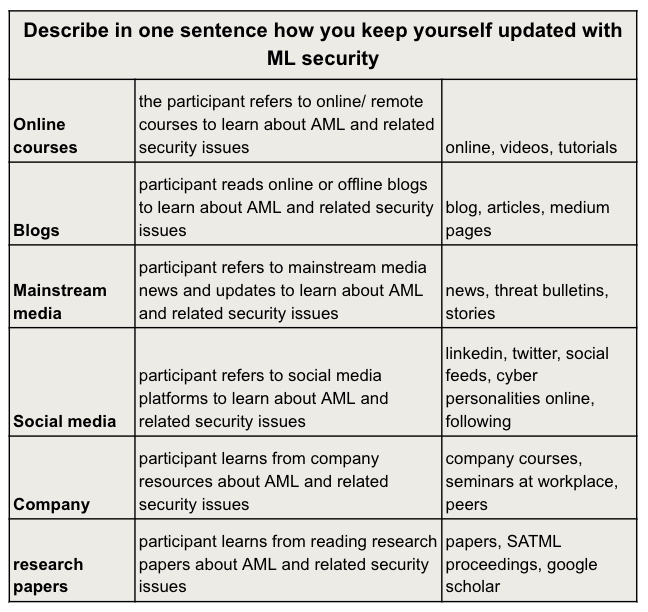}
  \vspace{-1em}
  \caption{Sources of education of AML threats}
  \label{fig:b}
\end{figure}

\begin{figure*}[t]
  \centering
  \includegraphics[width=0.75\linewidth]{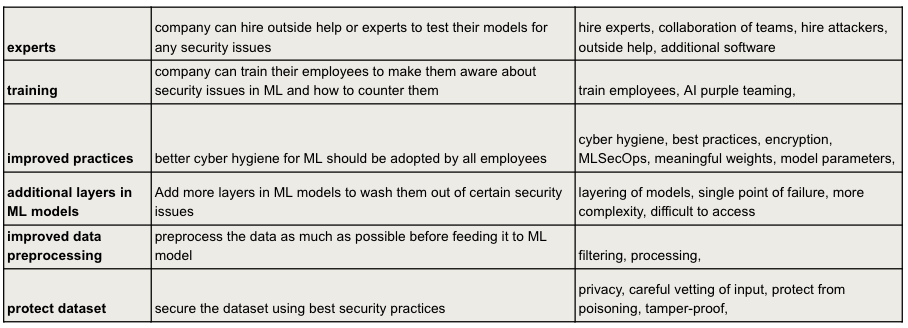}
  \vspace{-1em}
  \caption{Possible solutions for ML model protection for security}
  \label{fig:c}
\end{figure*}

\begin{figure*}[t]
  \centering
  \includegraphics[width=0.75\linewidth]{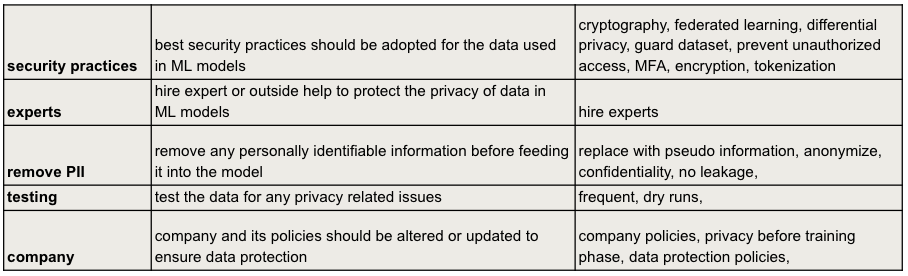}
  \vspace{-1em}
  \caption{Possible solutions for ML model protection for privacy}
  \label{fig:d}
\end{figure*}

As part of our Qualitative analysis, we created code-books based on the participant responses. They are shown in Figures~\ref{fig:a} through~\ref{fig:d}.

\subsection{Quantitative Analysis}
\label{sec:quantitative}

\subsubsection{First Study}

Statistical analysis of correlation between experience in cybersecurity and participants' belief regarding relevance of cybersecurity in ML. Here, F1=Extremely relevant, F2=Moderately relevant, F3=Relevant, F4=Not Relevant.

\begin{table}[h]
\caption{Fisher's exact test for Q1 and Q13}
\centering
\begin{tabular}{|l|c|c|c|c|c|}
\hline
\textbf{} & \textbf{F1} & \textbf{F2} & \textbf{F3} & \textbf{F4} & \textbf{Total} \\
\hline
No experience & 1 & 1 & 1 & 0 & 3 \\
Roughly one year & 1 & 1 & 0 & 0 & 2 \\
One to less than Five years & 0 & 3 & 0 & 0 & 3 \\
Five to less than Ten years & 1 & 1 & 0 & 0 & 2 \\
More than Ten years & 1 & 1 & 0 & 0 & 2 \\
\hline
\textbf{Column Total} & 4 & 7 & 1 & 0 & 12 \\
\hline
\end{tabular}

\end{table}

For Correlation factor between Q1 and Q13:
\[
\alpha = \frac{4! \cdot 7! \cdot 3! \cdot 2! \cdot 3! \cdot 2! \cdot 2!}{12! \cdot 3!} = 0.017
\]

\begin{table}[h]
\caption{Fisher's exact test for Q2 and Q13}
\centering
\begin{tabular}{|l|c|c|c|c|c|}
\hline
\textbf{} & \textbf{F1} & \textbf{F2} & \textbf{F3} & \textbf{F4} & \textbf{Total} \\
\hline
No experience & 1 & 1 & 0 & 0 & 2 \\
Roughly one year & 2 & 3 & 0 & 0 & 5 \\
One to less than Five years & 1 & 3 & 1 & 0 & 5 \\
Five to less than Ten years & 0 & 0 & 0 & 0 & 0 \\
More than Ten years & 0 & 0 & 0 & 0 & 0 \\
\hline
\textbf{Column Total} & 4 & 7 & 1 & 0 & 12 \\
\hline
\end{tabular}

\vspace{-2em}
\end{table}

For Correlation factor between Q2 and Q13:
\[
\alpha = \frac{4! \cdot 7! \cdot 3! \cdot 2! \cdot 5! \cdot 5!}{12! \cdot 3! \cdot 3! \cdot 2!} = 0.101
\]

\begin{table}[h]
\caption{Fisher's exact test for Q3 and Q13}
\centering
\begin{tabular}{|l|c|c|c|c|c|}
\hline
\textbf{} & \textbf{F1} & \textbf{F2} & \textbf{F3} & \textbf{F4} & \textbf{Total} \\
\hline
No experience & 1 & 3 & 1 & 0 & 5 \\
Roughly one year & 0 & 3 & 0 & 0 & 3 \\
One to less than Five years & 1 & 1 & 0 & 0 & 2 \\
Five to less than Ten years & 1 & 0 & 0 & 0 & 1 \\
More than Ten years & 1 & 0 & 0 & 0 & 1 \\
\hline
\textbf{Column Total} & 4 & 7 & 1 & 0 & 12 \\
\hline
\end{tabular}

\end{table}

For Correlation factor between Q3 and Q13:
\[
\alpha = \frac{4! \cdot 7! \cdot 5! \cdot 3! \cdot 2!}{12! \cdot 3! \cdot 3!} = 0.06
\]

\begin{table}[h]
\caption{Fisher's exact test for Q1 and Q20}
\centering
\begin{tabular}{|l|c|c|c|}
\hline
\textbf{} & \textbf{No} & \textbf{Yes} & \textbf{Row total} \\
\hline
No experience & 1 & 2 & 3 \\
Roughly one year & 1 & 1 & 2 \\
One to less than Five years & 3 & 0 & 3 \\
Five to less than Ten years & 0 & 2 & 2 \\
More than Ten years & 0 & 2 & 2 \\
\hline
\textbf{Column Total} & 5 & 7 & 12 \\
\hline
\end{tabular}

\end{table}

For Correlation factor between Q1 and Q20:
\[
\alpha = \frac{5! \cdot 7! \cdot 3! \cdot 2! \cdot 3! \cdot 2!}{12! \cdot 2! \cdot 2!} = 0.007
\]

\begin{table}[h]
\caption{Fisher's exact test for Q4 and Q13}
\centering
\begin{tabular}{|l|c|c|c|c|c|}
\hline
\textbf{} & \textbf{F1} & \textbf{F2} & \textbf{F3} & \textbf{F4} & \textbf{Total} \\
\hline
Yes & 1 & 6 & 0 & 0 & 7 \\
No & 3 & 1 & 1 & 0 & 5 \\
\hline
\textbf{Column Total} & 4 & 7 & 1 & 0 & 12 \\
\hline
\end{tabular}

\end{table}

For Correlation factor between Q4 and Q13:
\[
\alpha = \frac{4! \cdot 7! \cdot 7! \cdot 5!}{12! \cdot 6! \cdot 3!} = 0.035
\]

\begin{table}[h]
\caption{Fisher's exact test for Q6 and Q13}
\centering
\begin{tabular}{|l|c|c|c|c|c|}
\hline
\textbf{} & \textbf{F1} & \textbf{F2} & \textbf{F3} & \textbf{F4} & \textbf{Total} \\
\hline
Yes & 1 & 3 & 1 & 0 & 5 \\
No & 3 & 4 & 0 & 0 & 7 \\
\hline
\textbf{Column Total} & 4 & 7 & 1 & 0 & 12 \\
\hline
\end{tabular}

\end{table}

For Correlation factor between Q6 and Q13:
\[
\alpha = \frac{4! \cdot 7! \cdot 7! \cdot 5!}{12! \cdot 3! \cdot 3! \cdot 4!} = 0.001
\]

\begin{table}[h]
\caption{Fisher's exact test for Q8 and Q13}
\centering
\begin{tabular}{|l|c|c|c|c|c|}
\hline
\textbf{} & \textbf{F1} & \textbf{F2} & \textbf{F3} & \textbf{F4} & \textbf{Total} \\
\hline
Yes & 3 & 1 & 0 & 0 & 4 \\
No & 1 & 6 & 1 & 0 & 8 \\
\hline
\textbf{Column Total} & 4 & 7 & 1 & 0 & 12 \\
\hline
\end{tabular}

\end{table}

For Correlation factor between Q8 and Q13:
\[
\alpha = \frac{4! \cdot 7! \cdot 4! \cdot 8!}{12! \cdot 3! \cdot 6!} = 0.056
\]

\subsubsection{Second Study}

\begin{table}[h]
\caption{Fisher's exact test for Hypothesis - 1}
\centering
\begin{tabular}{|l|c|c|c|c|}
\hline
\textbf{} & \textbf{Yes} & \textbf{No} & \textbf{Neutral} & \textbf{Total} \\
\hline
Information Security & 5 & 0 & 3 & 8 \\
IoT & 1 & 0 & 0 & 1 \\
Computer Science & 3 & 0 & 0 & 3 \\
Electrical and Communication & 0 & 1 & 1 & 2 \\
Psychology & 0 & 1 & 0 & 1 \\
\hline
\textbf{Column Total} & 9 & 2 & 4 & 15 \\
\hline
\end{tabular}

\end{table}

The correlation factor between Q2 and Q28:
\[
\alpha = \frac{9! \cdot 2! \cdot 4! \cdot 8! \cdot 3! \cdot 2!}{15! \cdot 5! \cdot 3! \cdot 3!} = 0.006
\]

\begin{table}[h]
\caption{Fisher's exact test for Hypothesis - 2}
\centering
\begin{tabular}{|l|c|c|c|c|}
\hline
\textbf{} & \textbf{Yes} & \textbf{No} & \textbf{Neutral} & \textbf{Total} \\
\hline
Used CTF before & 8 & 2 & 4 & 14 \\
Not Used CTF before & 1 & 0 & 0 & 1 \\
\hline
\textbf{Column Total} & 9 & 2 & 4 & 15 \\
\hline
\end{tabular}

\end{table}

The correlation factor between Q7 and Q28:
\[
\alpha = \frac{9! \cdot 2! \cdot 4! \cdot 14!}{15! \cdot 8! \cdot 2! \cdot 4!} = 0.6
\]

\end{document}